\documentstyle[12pt]{article}
\begin{document}

\bibliographystyle{unsrt}    
\setcounter{secnumdepth}{0} 

\newcommand{\st}{\scriptstyle}
\newcommand{\sst}{\scriptscriptstyle}
\newcommand{\mco}{\multicolumn}
\newcommand{\epp}{\epsilon^{\prime}}
\newcommand{\vep}{\varepsilon}
\newcommand{\ra}{\rightarrow}
\newcommand{\ppg}{\pi^+\pi^-\gamma}
\newcommand{\vp}{{\bf p}}
\newcommand{\ko}{K^0}
\newcommand{\kb}{\bar{K^0}}
\newcommand{\al}{\alpha}
\newcommand{\ab}{\bar{\alpha}}
\def\be{\begin{equation}}
\def\ee{\end{equation}}
\def\bea{\begin{eqnarray}}
\def\eea{\end{eqnarray}}
\def\CPbar{\hbox{{\rm CP}\hskip-1.80em{/}}}

\def\ap#1#2#3   {{\em Ann. Phys. (NY)} {\bf#1} (#2) #3.}
\def\apj#1#2#3  {{\em Astrophys. J.} {\bf#1} (#2) #3.}
\def\apjl#1#2#3 {{\em Astrophys. J. Lett.} {\bf#1} (#2) #3.}
\def\app#1#2#3  {{\em Acta. Phys. Pol.} {\bf#1} (#2) #3.}
\def\ar#1#2#3   {{\em Ann. Rev. Nucl. Part. Sci.} {\bf#1} (#2) #3.}
\def\cpc#1#2#3  {{\em Computer Phys. Comm.} {\bf#1} (#2) #3.}
\def\err#1#2#3  {{\it Erratum} {\bf#1} (#2) #3.}
\def\ib#1#2#3   {{\it ibid.} {\bf#1} (#2) #3.}
\def\jmp#1#2#3  {{\em J. Math. Phys.} {\bf#1} (#2) #3.}
\def\ijmp#1#2#3 {{\em Int. J. Mod. Phys.} {\bf#1} (#2) #3.}
\def\jetp#1#2#3 {{\em JETP Lett.} {\bf#1} (#2) #3.}
\def\jpg#1#2#3  {{\em J. Phys. G.} {\bf#1} (#2) #3.}
\def\mpl#1#2#3  {{\em Mod. Phys. Lett.} {\bf#1} (#2) #3.}
\def\nat#1#2#3  {{\em Nature (London)} {\bf#1} (#2) #3.}
\def\nc#1#2#3   {{\em Nuovo Cim.} {\bf#1} (#2) #3.}
\def\nim#1#2#3  {{\em Nucl. Instr. Meth.} {\bf#1} (#2) #3.}
\def\np#1#2#3   {{\em Nucl. Phys.} {\bf#1} (#2) #3.}
\def\pcps#1#2#3 {{\em Proc. Cam. Phil. Soc.} {\bf#1} (#2) #3.}
\def\pl#1#2#3   {{\em Phys. Lett.} {\bf#1} (#2) #3.}
\def\prep#1#2#3 {{\em Phys. Rep.} {\bf#1} (#2) #3.}
\def\prev#1#2#3 {{\em Phys. Rev.} {\bf#1} (#2) #3.}
\def\prl#1#2#3  {{\em Phys. Rev. Lett.} {\bf#1} (#2) #3.}
\def\prs#1#2#3  {{\em Proc. Roy. Soc.} {\bf#1} (#2) #3.}
\def\ptp#1#2#3  {{\em Prog. Th. Phys.} {\bf#1} (#2) #3.}
\def\ps#1#2#3   {{\em Physica Scripta} {\bf#1} (#2) #3.}
\def\rmp#1#2#3  {{\em Rev. Mod. Phys.} {\bf#1} (#2) #3.}
\def\rpp#1#2#3  {{\em Rep. Prog. Phys.} {\bf#1} (#2) #3.}
\def\sjnp#1#2#3 {{\em Sov. J. Nucl. Phys.} {\bf#1} (#2) #3.}
\def\spj#1#2#3  {{\em Sov. Phys. JEPT} {\bf#1} (#2) #3.}
\def\spu#1#2#3  {{\em Sov. Phys.-Usp.} {\bf#1} (#2) #3.}
\def\zp#1#2#3   {{\em Zeit. Phys.} {\bf#1} (#2) #3.}

\setcounter{secnumdepth}{2} 

\rightline{\tenrm Brown-HET-1024}
\rightline{\tenrm TA-530}
\rightline{\tenrm McGill/95--57}

{}~\bigskip

{}~\bigskip
\centerline{\bf A REVIEW OF THE SOFT POMERON$^a$}
{}~\\
{}~\\
\centerline{J.R. Cudell$^b$ }
\centerline{\tenrm Physics Department, Brown University$^c$, Providence,
RI 02906, U.S.A.}
\centerline{\tenrm and
Physics Dept., McGill University, Montr\'eal,
Qu\'ebec, H3A 2T8, Canada}
{}~\\

\begin{quote}{\centerline{\tenbf Abstract}
\tenrm Soft pomeron fits reproduce all
zero-$Q^2$ data for light quarks, but run into problems at HERA
for heavy mesons and for high $Q^2$, and at the Tevatron for $W$ diffractive
production. I review the basic properties of the soft pomeron, and outline the
possibilities which have been considered to account for the new data. }
\end{quote}
\section{Light quarks at $Q^2=0$:}
\renewcommand{\thefootnote}{}
At
\footnote{$^a$ Invited talk at the International
Europhysics Conference on High Energy
Physics (HEP 95),
27 Jul - 2 Aug 1995 , Brussels, Belgium}
high energy, pomeron exchange\footnote{$^b$Permanent address:
Institut de Physique, Universit\'e de Li\`ege, Sart Tilman,  B-4000 Li\`ege,
Belgium, E-Mail: cudell@gw.unipc.ulg.ac.be.}
controls\footnote{$^c$Supported in part by USDOE contract DE-FG02-91ER
40688-Task A.}
 total cross sections, single- and double- diffractive cross
sections and elastic cross sections.  All these rise very slowly with $s$, and
whatever is responsible for that rise is called the soft pomeron. We shall
examine only two of the most extreme models~\cite{DL,Capella} which
successfully reproduce the data. All existing
models differ by their answers to the following questions:

\noindent$\bullet$ {\it Is the
pomeron a Regge trajectory?} One of the most successful models is that of
Donnachie and Landshoff (DL)~\cite{DL}, who fit all the soft cross
sections using two Regge trajectories: a degenerate one for the $\rho$ (of
charge parity $C=-1$) and $a$ ($C=+1$) exchanges, and another for the
pomeron. As the center-of-mass energy
$\sqrt{s}$ becomes large, and for momentum transfers $t\leq 1$ GeV$^2$,
 the pomeron trajectory dominates
the hadronic elastic amplitude, which behaves like $s^{1.08+0.25 t}$.
{}From such a trajectory, Regge theory predicts the existence of a $2^{++}$
glueball, with a mass of the order of 1.9 GeV.  A glueball candidate has
been found precisely at that mass~\cite{WA}.

Note that a simple-pole behaviour is not required to fit the slow increase
of the soft cross sections: several other
models reproduce it very successfully, and the basic
difference leads to the second issue.

\noindent $\bullet${\it Unitarisation:} A simple-pole behaviour
will eventually break the Froissard bound. Hence, once an amplitude
increasing with $s$ faster than a $\log^2 s$ is assumed, one needs to
confront the question of unitarisation.
As one deals with high-s, eikonal
methods are applicable: one can go to impact parameter
space, calculate multiple exchanges there, then sum them and go back to $t$.

Nevertheless, we do not know what
the pomeron couples to.
If it couples to quarks, then two-pomeron exchange contains
more diagrams than the standard eikonal expansion, as
multiple exchanges can couple to different quarks. In the DL model~\cite{DL},
at the two-pomeron-exchange level, this leads to
a substantial reduction of the coefficient of the eikonal expansion: thus
the effect of unitarisation is weak, and brings the
one-pomeron intercept to~1.085.

However, the proton can clearly
fluctuate into non-elastic states, which eventually recombine into a proton.
In an extended eikonal formalism,\break
 Capella, Kaidalov, Merino, Pertermann and Tran
Thanh Van (CKMPT)~\cite{Capella} find
that such fluctuations increase the effect
of unitarisation. This enables them to accommodate a larger
one-pomeron exchange intercept, of the order of 1.28.

Finally,  one may note
that it is possible to fit the data with only $\log s$ and $\log^2 s$
terms~\cite{Kang}, and hence assume that only the unitarised
asymptotic amplitudes matter.
It is clear however that some properties of the soft
exchanges are lost if unitarisation is strong.

\noindent $\bullet${\it Factorisation
and quark counting:}  One of the main reasons to assume that the pomeron
couples to valence quarks is the quark counting rule, which seems to work
for pion-proton cross sections, as well as for cross sections involving
strange quarks~\cite{DL}. Hence it seems that quark degrees of freedom are
relevant for soft cross sections. Similarly, the amplitudes
factorise~\cite{factorisation} into one factor associated with the target,
one factor associated with the projectile, and a third factor describing the
exchange. This implies that the pomeron couples to one quark at a time:
otherwise, the exchange would feel the hadronic wave-function, and one would
have a convolution that does not factorise. These two properties are
violated by perturbative QCD, as well as by strong unitarisation.
Nevertheless, our
ignorance of hadronic wavefunctions can easily
accommodate the existing data, but
one has then to assume that the wavefunction of a pion is similar to that
of a proton, contrarily to the simplest intuition.

The DL model has received striking confirmation from HERA. The total
$\gamma p$ cross section was exactly predicted~\cite{DL}, and the most
recent data~\cite{HERAtotal} confirm this fact. Furthermore, one can
also calculate photoproduction cross sections: one needs to invent a vertex
describing the conversion of a photon into a vector meson, and this vertex
controls the magnitude of the cross section. On the other hand, the energy
dependence of the cross section entirely comes from the exchanged
trajectories ($a$ and pomeron). The DL model can be applied to
photoproduction~\cite{vector}, and works perfectly for $\rho^0$
photoproduction~\cite{HERAvector,NMC}.  There are violations of the
quark counting rule for $\Phi$, although the observed energy dependence is
consistent with a simple-pole parametrisation.

In the case of J/$\psi$
photoproduction~\cite{HERAvector}, it seems however that one encounters the
first violation of the simple-pole approach: the cross section rises as
$s^{0.4}$.  One might question the experimental results, as the
prediction holds only in the case of elastic scattering, {\it i.e.}
$\gamma^* p\rightarrow \rho^0 p$, without any break-up of the proton. As the
proton is not seen, it is hard to assert what the exact background is, and
some contamination is possible from $\gamma^* p\rightarrow \rho^0 X$.
Future runs at HERA will tell us if this is a real effect, as the
forward proton has become detectable in the upgraded detectors.
\section{Deep Inelastic Scattering} At high $Q^2$, the simple-pole picture
of the soft pomeron seems to break down. There are two instances in which
this is manifest.

\noindent $\bullet$ {\it Structure functions:}  At small $x$,
$F_2$ is expected to be dominated by pomeron exchange~\cite{protonsf}: as
$x\rightarrow 0$ the
sub-energy flowing through $F_2$ is $s={k^2+k_T^2\over x}$, with $k^2$ the
virtuality of the struck quark, and $k_T$ its transverse momentum.
Hence at small $x$ the sub-process enters a kinematic range close to
that of total cross sections. However, it seems that $Q^2\neq 0$ introduces
a drastic change in behaviour, even for $Q^2$ as low as 2 GeV$^2$: the
``effective intercept'' is of the order of $1.35$~\cite{HERASF}.  Such a
growth of the intercept comes naturally in the CKMPT model~\cite{Capella},
 as unitarisation
weakens when $Q^2$ increases.

\noindent $\bullet${\it Vector Meson Production}
The same disagreement can be found in
DIS vector meson production~\cite{HERAvector}. Up to NMC energies,
deep-inelastic production of vector mesons ($\rho^0$ and $\Phi$) is well
reproduced by simple-pole models~\cite{vector}, even at high $Q^2\sim \rm 25~
GeV^2$. At HERA, however, DIS production of vector mesons seems to follow
the same trend as $F_2$, and the effective intercept does not seem to be
$Q^2$-dependent.

As there are now several models on the market, it might
be worth pointing out in what way they differ. The first one~\cite{vector}
was introduced by Donnachie and Landshoff, who also realised the analogy with
lowest-order QCD. This analogy was refined by the present author~\cite{JRC},
and remarkable agreement has been obtained both with EMC and NMC
data~\cite{NMC}. At lowest-order, the relation between lowest-order $\rho^0$
production and the gluon structure function is obvious~\cite{JRC2}.
Ryskin, and later Brodsky et al.,~\cite{Ryskin etc}
 have argued that this relation extends to higher
orders of perturbation theory. One must realise that this often-quoted result
 is only
approximate: it is a perturbative, leading-log result, which holds
at $t=0$, and which is supposed to relate two non-perturbative quantities.
Unfortunately, $t$ cannot be zero in this process, and there has always to
be a non-negligible longitudinal momentum transferred to the photon in order
for it to convert into a $\rho$. Hence the equivalent of Bjorken-$x$ is
ill-defined, and the relation can only be approximate.

All the above models agree on some of the
predictions, namely that the cross section should be dominated by its
longitudinal part at large $Q^2$, and that it should behave like $1/Q^6$
(there might be room for a log~$Q^2$ in the perturbative case). The data
from Zeus gives $1/Q^{\left(4.2\pm 0.4^{+1.4}_{-0.5}\right)}$.
 \section{The pomeron structure function}
One way to inquire about the partonic content of the pomeron
is to probe it directly. UA8 has shown that hard
diffraction does exist~\cite{UA8} and Donnachie and Landshoff
predicted a long time ago that some 10\% of the hard events at
HERA would contain a rapidity gap~\cite{DLhard}. In this kind
of model~\cite{IngelmanSchlein}, the cross section is split
into a ``pomeron flux factor'' and ``pomeron structure
function''.

 The first factor comes from the coupling of the
pomeron to protons, and the second factor is the analog
 of the usual proton $F_2$. A word of caution is in order: as
the mass shell of the pomeron is far away, the cross section cannot be
continued to a region where the pomeron is on-shell, {\it i.e.} where its
flux is well-defined. Hence the splitting between structure function and
flux factor is
defined up to a factor, and there is no momentum sum rule for the
pomeron structure function. The pomeron flux factor however
must behave like
$x_P^{-1+2<\alpha(t)>}$, with $x_P$ the fraction of
longitudinal momentum carried by the pomeron, and
$<\alpha(t)>$ the average pomeron intercept over the $t$-range
of the experiment. For a soft pomeron, one expects~\cite{stirling} $-1+2
<\alpha(t)>=1.11 \pm 0.03$. The HERA data is compatible with this,
but is also compatible with a harder behaviour~\cite{pomsf}.

The pomeron structure function has also been measured~\cite{pomsf}.
It is worth noting
that although such an object can be empirically defined, it is not clear
that it is universal, and that the same structure function holds in
different processes~\cite{pomsfdef}. From the DL model,
one expects a quark structure function $\beta q(\beta)\approx 0.2 \beta
(1-\beta)$. There are several determinations which tend to the conclusion
that the pomeron is mostly hard, and that it is made of $30-80\%$
gluons~\cite{UA8,pomsf}, with a ``superhard'' component~\cite{superhard},
where the pomeron carries all the momentum of the hadron, present in
hadronic data.

There is however one major
problem: whatever the soft pomeron is made of, it is couples to
quarks, and hence its quark structure function cannot be zero-HERA
has actually measured it. This seems in
blatant contradiction with recent data from CDF~\cite{CDFpom},
who fail to detect $W$
production within a rapidity gap, when a few percent of the total production
would be expected in the above picture. This has prompted
Goulianos~\cite{goulianos} to renormalise the pomeron flux. A less drastic
method
would be to unitarise the pomeron, and hence to decrease its flux at high
energies only. This problem needs to be addressed before meaningful predictions
can be made for the LHC.
 \section{Conclusion}
A simple-pole model works perfectly for $x\geq 0.01$, and
up to $Q^2\approx 25$
GeV$^2$. At smaller x values, there are problems for large
 quark masses, and for high $Q^2$.

No model can explain both these facts.
Strong unitarisation \`a la\break CKMPT~\cite{Capella} can succeed in fitting
the HERA data,
but abandons factorisation and quark counting, and cannot account for
high-$Q^2$ data at moderate $s$. This model predicts that there is a
maximum intercept of the order of 1.3. It is of course tempting to see the
rise of a perturbative component~\cite{BFKL} in the data. However, at present,
 models mixing a hard pomeron and a
soft one~\cite{hardsoft} do not predict the $Q^2$ dependence of the data,
although they predict an intercept varying with $s$, with a maximum
intercept much larger, of the order of 2.

HERA and CDF data point
to the importance of unitarisation, and to the
multi-component nature of the pomeron, but we are still far from
understanding how to consistently build a model for this.

High-$t$ data will tell us a lot, as
they might exhibit the same kind of violation as high-mass and high-$Q^2$
data. They will directly probe the trajectory of the pomeron, which for
glueball exchange should be linear. Besides, the soft pomeron tends to
produce distributions which fall rapidly with $t$, whereas pQCD predicts a
much slower fall-off at high $t$: a weaker $t$ dependence in elastic events
would be a signature for a hard pomeron.

Finally, factorization and quark counting are essential to our
understanding of the pomeron, and can be tested at HERA, {\it e.g.} if
factorisation holds, the ratio of the diffractive cross section over the
elastic one should be the same for J/$\psi$ and for $\rho_0$, and should be
the same at different $Q^2$. Hence we urge the experimentalists not to
throw away their diffractive ``background'' events!
\setcounter{secnumdepth}{0}
\section{Acknowledgements}
I thank K. Kang and P.V. Landshoff for discussions, and C.A. Pont for a careful
proofreading.
  \end{document}